# Nuclear lattice model and the electronic configuration of the chemical elements

Jozsef Garai

*Department of Civil Engineering, University of Debrecen, Hungary*
*E-mail address: jozsef.garai@fiu.edu*

**Abstract**  The fundamental organizing principle resulting in the periodic table is the nuclear charge. Arranging the chemical elements in an increasing atomic number order, a symmetry pattern known as the Periodic Table is detectable. The correlation between nuclear charge and the Periodic System of the Chemical Elements indicates that the symmetry emerges from the nucleus. Nuclear symmetry can only exist if the relative positions of the nucleons in the nucleus are invariant. Pauli exclusion principle can also be interpreted as the nucleons should occupy a lattice position.  Based on symmetry and other indicatives face centered cubic arrangement have been proposed for the nuclear lattice.  The face centered cubic arrangement consists exclusively tetrahedron and octahedron sub-units.  Based on the higher density of the tetrahedron sub-unit, the length of the first period and other considerations it is suggested that the nuclear structure should be developed from a tetrahedron "seed".  Expanding this tetrahedron sub-unit and assuming that protons and neutrons are alternate arranged then the number of the protons in the expanded layer of the tetrahedron are identical with the main quantum numbers. The reproduction of the angular and the magnetic momentum quantum numbers by the lattice positions are also identical.  Denser nuclear structure  can be achieved if the expansion of the tetrahedron is shifted by 90 degrees rotation at each step.  The number of protons in the layers of this "double" tetrahedron reproduces the periodicity of the chemical elements. Mathematical equations for all of the sequences of the periodic table are derived by counting the nucleons in the tetrahedron  nuclear model.  One of the important outcome of the lattice structure of the nucleus is that the nucleus can not be considered as a point charge because the position of the individual protons are varies and their attraction on the electrons are different.  Following classical approach it can be predicted that as long as the attraction of an added proton to the structure is stronger than the existing ones then the captured electron joins to the valence shell of the electrons.  This condition practically is the reiteration of the Aufbau principle.  When the attraction on the captured electron is weaker than on the existing ones then a new valence electron shell should be started.  The relative strength of the attraction of the protons is described by the distance between the proton and the charge center of the nucleus.  Shorter distance corresponds to weaker attraction and vice versa.  Based on the geometry of the nuclear structure it is shown that when a new "layer" of the nuclear structure starts then the distance between the first proton in the new layer and the charge center of the nucleus is smaller than the distance of the proton, which completed the preceding "layer".  Thus a new valence electron shell should start to develop when the nuclear structure is expanded.  The expansion of the double tetrahedron

FCC nuclear lattice model offers a feasible physical explanation how the nucleus affects the electronic configuration of the chemical elements depicted by the periodic table.

**Keywords**   Periodic System of the Chemical Elements  ▪  Periodic table  ▪  Electronic Structure of the Elements  ▪  Nuclear Structure  ▪  Nuclear Lattice Model  ▪  Sequences of the Periodic Table  ▪  Aufbau Principle  ▪  Pauli Exclusion Principle

**1. Introduction**

*1.1 Preface*

In the earliest days of science researchers were arguing philosophically what was the reasonable reason for an observed phenomenon.  The majority of the contemporary scientific community claim that these arguments are useless because they do not add anything to our understanding of nature.  The current consensus on the aim of science is that science collect facts (data) and discern the order that exist between and amongst the various facts (e.g. Feynman, 1985).  According to this approach the mission of science is over when the phenomenon under investigation has been described.  It is left to the philosophers to answer the question why and figure it out what is the governing physical process of the phenomenon.  Quantum mechanics is a good example of this approach "It works, so we just have to accept it".  The consequence is that nearly 90 years after the development of quantum theory, there is still no consensus in the scientific community regarding the interpretation of the theory's foundational building blocks (Schlosshauer et al. 2013). I believe that identifying the physical process governing a natural phenomenon is the responsibility of science.  Dutailly (2013) expressed this quite well: A "black box" in the "cloud" which answers rightly to our questions is not a scientific theory, if we have no knowledge of the basis upon which it has been designed. A scientific theory should provide a set of concepts and a formalism which can be easily and indisputably understood and used by the workers in the field.

   In this study the main unifying principle in chemistry, the periodic system of the chemical elements (PSCE) is investigated.  The aim of the study is not only the description of the periodicity but also the understanding of the underlying physics resulting in PSCE.

*1.2. The periodic pattern of the chemical elements emerged*

By 1860 about 60 elements were identified, and this initiated a quest to find their systematic arrangement.  Based on similarities, Döbereiner in 1829 suggested grouping the elements into triads.  Newlands (1864) in England arranged the elements in order of increasing atomic weights and, based on the repetition of chemical properties, proposed the "Law of Octaves".  Listing the elements also by mass, Mendeleev (1869; 1872) in Russia and Meyer (1870) in Germany simultaneously proposed a 17-column arrangement with two partial periods of seven elements each (Li-F and Na-Cl) followed by nearly complete periods (K-Br and Rb-I). Mendeleev gets higher credit for this discovery because he published the results first; he also rearranged a few elements out of strict mass sequence in order to fit better to the properties of their neighbors and corrected mistakes in the values of several atomic masses. Additionally, he predicted the existence and the properties of a few new elements by leaving empty cells in his table. Mendeleev's periodic table did not include the noble gasses, which were discovered later. Argon was identified by Rayleigh in 1895/a-b. The remainder of the noble gasses were discovered by Ramsey (1897) who positioned them in the periodic table in a new column. Anton van den Broek (1911; 1913) suggested that the fundamental organizing principle of the table is not the weight but rather the nuclear charge, which is equivalent with the atomic number. The extended 18-column table was slightly modified based on Moseley's experiments (1913), and he rearranged the table according to atomic number. The discovery of the transuranium elements from 94-102 by Seaborg (1951) expanded the table. He also reconfigured the table by placing the lanthanide/actinide series at the bottom. There is no "standard" or approved periodic table. The only specific recommendation of the International Union of Pure and Applied Chemistry (IUPAC), which is the governing body in Chemistry, is that PSCE should follow the 1 to 18 group numbering (Leigh, 2009) (Fig. 1).

*1.3. Sequences of the Table*

The organizing pattern of PSCE can be described by three sequences. These digital descriptions of the table are the fundamental [$S_{fundamental}$], the periodic [$S_{\Delta Z}$], and the atomic number [$S_Z$] sequences (Fig. 1) (Garai, 2008). The fundamental sequence of the table is:

$$S_{fundamental} = \left\{ 1, 2, 2, 3, 3, 4, 4, \ldots \right\} \qquad [1]$$

The number of elements [$\Delta Z(n)$] within the period or the length of the periods has the sequence of

$$S_{\Delta Z} = \left\{ 2, 8, 8, 18, 18, 32, 32 \ldots \right\}. \qquad [2]$$

The atomic number or the nuclear charge of the elements [$Z(n)$] in a completely developed period follows the sequence

$$S_Z = \left\{ 2, 10, 18, 36, 54, 86, 118 \ldots \right\}. \qquad [3]$$

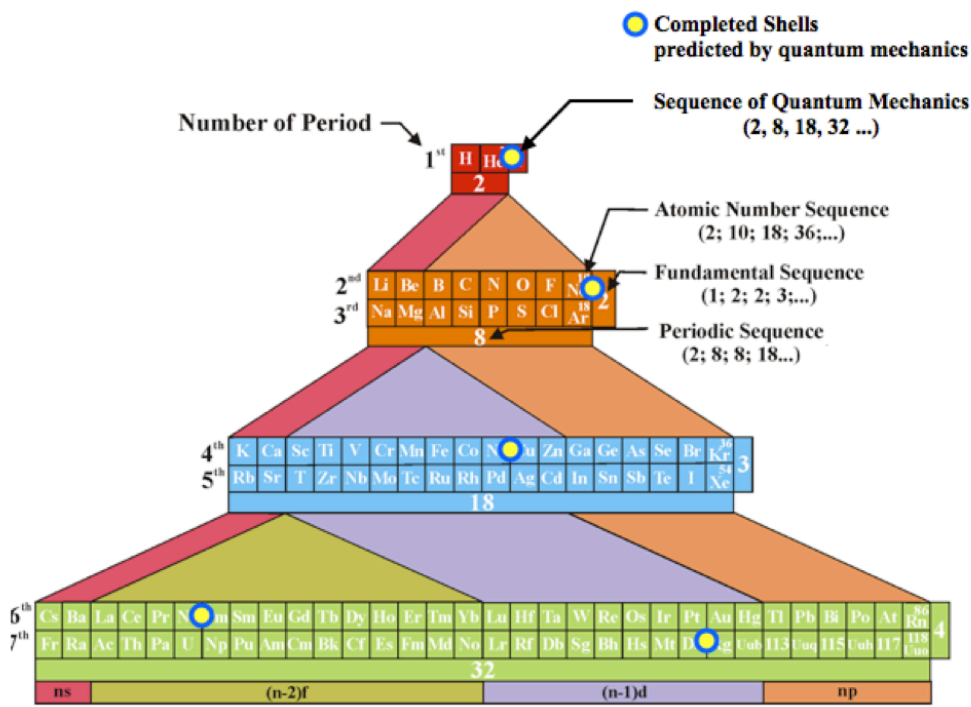

**Fig. 1.** The sequences of the periodic table. The sequence derived from quantum theory is also shown. It can be seen that the sequence derived from quantum theory does not match with the sequences of PSCE, indicating that the theory should not be considered as a viable physical explanation for the periodicity

## *1.4 Pauli Exclusion Principle*

In order to explain the observed light emission patterns observed in atoms Pauli postulated the exclusion principle: "In an atom there cannot be two or more equivalent electrons for which the values of all four quantum numbers coincide. If an electron exist in an atom for which all of these numbers have defined values, then this state is occupied." (Pauli, 1925; 1964). The principle originally was postulated to electrons but it has been generalized and now includes other particles. The Pauli Exclusion Principle is one of the corner stones of quantum physics and thus it is at the basis of the foundation of modern physics. It is connected to spin statistics dividing the world into fermions and bosons. Particles with half-integer spin (fermions) are described by antisymmetric wave functions and particles with integer spin (bosons) are described by symmetric wave function (Pauli, 1946). The rise of the periodic table is one of the important outcome of the Pauli exclusion principle.

Pauli in his Nobel lecture in 1946 (Pauli, 1946) stated that "I was unable to give a logical reason for the exclusion principle or to deduce it from more general assumptions. I had always the feeling and I still have it today, that this is a deficiency." After many decades past the physical explanation for Pauli's postulate is still lacking (Kaplan, 2013). Pauli exclusion principle is the theoretical basis for the periodic table.  Thus the physical explanation for the periodicity of the chemical elements is also lacking.

*1.5 Physical Explanations*

The Aufbau principle, how the electronic configuration of the atoms built up, originally was postulated by Niels Bohr and Wolfgang Pauli.  They stated that "The orbitals of lower energy are filled in first with the electrons and only then the orbitals of high energy are filled.".  The energy levels calculated from the hydrogenic model, or the main quantum number follows the sequence of $2n^2$.  Only the first two entries of this quantum number sequence is consistent with the periodic sequence.  In order to explain the periodic sequence by quantum theory the (n, l) rule or the Madelung energy ordering (Karapetoff, 1930; Madelung, 1936) has been suggested, which applies to neutral atoms in their ground state.  The majority of general and physical chemistry books present the symmetry of PSCE as satisfactorily explained by either the electronic structure of the elements (Bohr, 1922; Hund, 1925; Slater, 1930; Condon, 1935; Landau & Lifschitz, 1977; Schwabl, 2001; Atkins & Atkins, 2001) or by quantum mechanics (Hartree, 1957; Fischer, 1977; Johnson, 2005).  Some authors present quantum justification of the rule (Demkov and Ostrovsky, 1971; Ostrovsky, 1981, 2001).  Not all of the elements comply with the (n,l) rule; therefore, general acceptance is lacking (Scerri 2004; Boeyens, 2008).  Based on the conflicting interpretations it can be concluded that the most important periodic sequence of the chemical elements has not been satisfactorily explained (Scerri, 1998; Schwartz and Wang, 2010; Boeyens, 2013).  One example might be the positions of Hydrogen and Helium.  Hydrogen has one 1s electron but also one electron is needed to attain inert configuration.  Thus it can be either placed in the 1st group or in the 17th group.  Based on chemical behavior, Hydrogen neither halogen nor alkali metal but rather both.  Thus the position of Hydrogen in the PSCE is uncertain (Scerri, 2007).  Helium with its $1s^2$ electron configuration is the other element with uncertain position.  Helium should be in the 2nd group.  However, based on its chemical properties, equivalent to inert gas, is placed into group 18.  Besides Helium, the outermost electron configuration of group 18th is $2p^6$.  No theoretical explanation for the shift in the electron configuration from $2p^6$ to $1s^2$ or vice versa is offered.  Based on these deficiencies, it has been suggested that quantum mechanics is unable to explain the most important aspects of the periodic table (Scerri, 1998; Boeyens, 2008; Schwartz and Wang, 2010; Boeyens, 2013). The substantial number of articles in the current literature (e.g. Sneath, 2000; Giunta, 2001; Kragh, 2001; Ostrovsky, 2001; Scerri 2001; Dudek et al. 2002; Baum, 2003; Dordrecht, 2003; Ostrovsky, 2003; Moore, 2003; Scerri, 2003; Friedrich, 2004; Kibler, 2004;  Schunck & Dudek, 2004; Scerri, 2004; Schwarz, 2004; Rouvray & King, 2005; Bent, 2006; Rouvray & King, 2006 ; Wang, 2006; Restrepo & Pachon 2007; Schwarz, 2007; Weinhold & Bent, 2007; Boeyens & Levendis, 2008; Wang & Schwartz, 2009) discussing the fundamental problems of PSCE nearly one-and-a-half centuries after its invention, also indicates that the complete physical understanding of the symmetry expressed by PSCE is still lacking.

Despite the remaining outstanding questions, there is a general agreement that the elements should be arranged in an increasing atomic number order in PSCE. Invariant symmetry relatig to the nuclear charge can only be possible if the positions of the nucleons are invariant too.

2. **Nuclear Models**

The most widely accepted models for the nucleus are the shell (Mayer, 1949; Haxel et al., 1949; Rainwater, 1950; Mayer & Jensen, 1955), the liquid drop (Bohr, 1936; Feenberg, 1947) and the cluster (Hafstad & Teller, 1938) models. The shell model assumes a gas phase (Fermi) for the nucleus and is able to explain the independent quantum characteristics of the nucleons. The liquid drop model is able to explain the observed saturation properties of the nuclear forces, the low compressibility of the nucleus, the well-defined nuclear surface, the binding energies and most importantly the fission phenomena. The clustering of alpha particle model was deduced from the fact that certain large nuclei emits alpha particles and the stability and the abundance of the 4n-nuclei is significantly higher. These models are able to describe successfully certain selected properties of the nucleus; however, none of them able to give a comprehensive description. The basic assumptions of the models are contradictory; therefore, it is impossible combine them and develop a hybrid model.

The assumed phase of the nucleus in these models is gas, liquid or semi-solid. None of these phases have an invariant nucleon position; therefore, these models cannot maintain the symmetry present in the PSCE. The preservation of symmetry requires a "solid phase" nucleus, in which the positions of the nucleons are preserved. "Solid phase" or lattice models have not been considered for many decades as a viable option because of the uncertainty principle and the lack of diffraction. In the 1960s, the discovery of quarks and neutron star research satisfactorily answered these objections and opened the door for lattice models. The first nuclear lattice models were presented by Linus Pauling in 1965/a-c.

The lattice models can easily reproduce the various shell, liquid-drop, and cluster properties (Cook & Dallacasa, 1987; Cook & Hayashi, 1997). Asymmetric fission and heavy-ion multi-fragmentation are some of the phenomena that the traditional models of nuclear structure theory cannot explain, yet can be reproduced by lattice models (Gupta et al., 1996; 1997).

Significant effort has been made to find correlation between lattice positions and quantum numbers with partial success for FCC structure (Pauling, 1965/d; Cook & Dallacasa, 1987). The symmetries of the Schrödinger's equation are also correspond to FCC geometry (Wigner, 1937). The common features of the developed lattice models are that the protons and the neutrons have the same size and they are alternately arranged in a closest packing array (Anagnostatos,1973; Canuto & Chitre, 1974; Lezuo, 1974; Cook, 1976; Matsui et al., 1980; Dallacasa, 1981). These assumptions are reasonable.

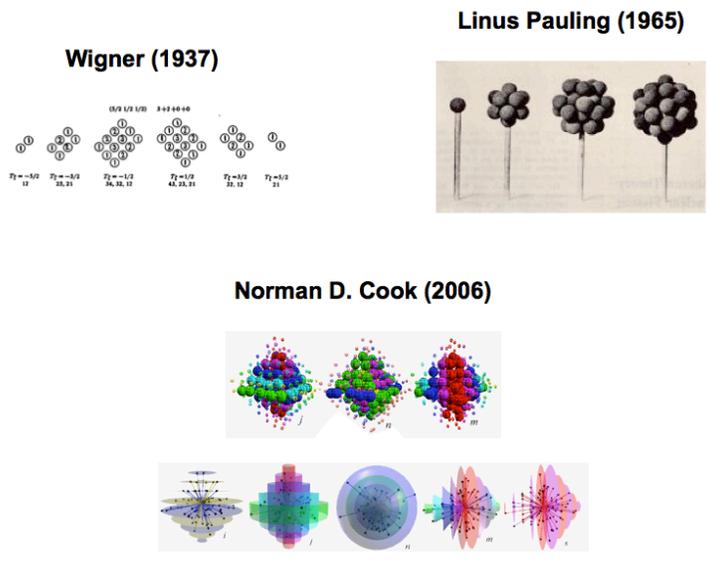

**Fig. 2.** Previous lattice models of the nucleus. These investigations expanded the FCC nuclear lattice structure spherically.

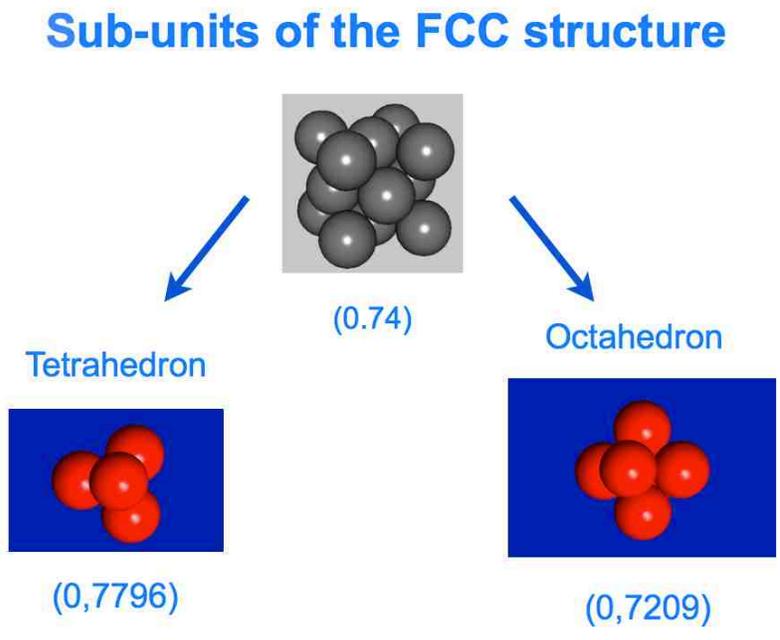

**Fig. 3.** The sub-units of the face centered cubic structure and their densities. The density of the FCC structure is 0.74. The density of the sub-units, tetrahedron and octahedron are 0.7796 and 0.7209 respectively. The ratio of the tetrahedron and octahedron sub-units in the FCC structure is 2:1.

The radii of protons and the neutrons differ only slightly (Schery et al., 1980). The same proton and neutron magic numbers indicate same structural development for both protons and neutrons, which is consistent with an alternate arrangement. The equal spheres will most likely utilize the available space in the most efficient way, which is a closest packing arrangement. Previous investigations (Pauling, 1965/d; Cook & Dallacasa, 1987), which expanded the FCC structure spherically, had partial success in finding correspondence between lattice positions and quantum states (Fig. 2). The structure of the closest packing arrangements consists of tetrahedron and octahedron sub-units exclusively. The expansion of these sub-units should be investigated first if someone wants to look for structural symmetry patterns in the FCC lattice (Garai, 1999; 2003) (Fig. 3).

## 3. Tetrahedron FCC Lattice Model

Assuming that the bonding energy correlates to the nucleon density then the densest structure should be formed preferably. The densities of the tetrahedron and octahedron units in FCC are 0.7796 and 0.7209 respectively (Garai, 2010) indicating preferential tetrahedron formation.

The first sequence of the nuclear structural development is completed by the nucleus of Helium, which contains four nucleons. The closest packing arrangement of the four nucleons is tetrahedron. Forming a tetrahedron sub-unit nucleus in the first completed period is an additional support for tetrahedron nucleus formation. Calculations of potential models, constrained by the hadron spectrum for the confinement of the relativistic quark (Goldman et al. 1988; Maltman et al. 1994) and colored quark exchange model (Robson, 1978), are also consistent with a tetrahedron forming the He nucleus. The expansion of this tetrahedron seed of four nucleons is investigated here.

The tetrahedron shape of equal spheres arranged in FCC packing can be formed from layers of equilateral triangles packed in two dimensional closest packing arrangement as shown in Fig. 4/a. Starting with one sphere and increasing the length of the side of the triangles by one additional sphere, the number of nucleons in each triangle plane will be 1, 3, 6, 10, 15, 21, 28, 36... (Fig. 4/a). Stacking these layers, the number of spheres in two consecutive layers are 4, 16, 36, 54... Assuming that protons and neutrons are alternately arranged in the lattice the number of spheres should be divided by two. This gives the proton numbers to 2, 8, 18, 32... respectively (Fig. 4/b). These numbers are identical with the number of possible states of the principle quantum numbers. If the tetrahedron is expended on two faces then a shell-like structure can be formed (Fig. 4/c), which is consistent with the physical interpretation of the principle quantum number.

Investigating how many spheres can be accommodated in one row in the outer shell gives the total number of spheres in one row to 4, 12, 20, 28..., which corresponds to 2, 6, 10, 14... proton number (Fig. 5/a). These proton numbers are identical with the number of states determined by the angular momentum quantum numbers corresponding to s, p, d and f orbitals. The rows in the outer layers of the tetrahedron are one unit distance away from each other; thus the identical agreement of the number of nucleons with the angular momentum quantum numbers is not only in numerical agreement but also bears the same physical meaning defined by quantum theory.

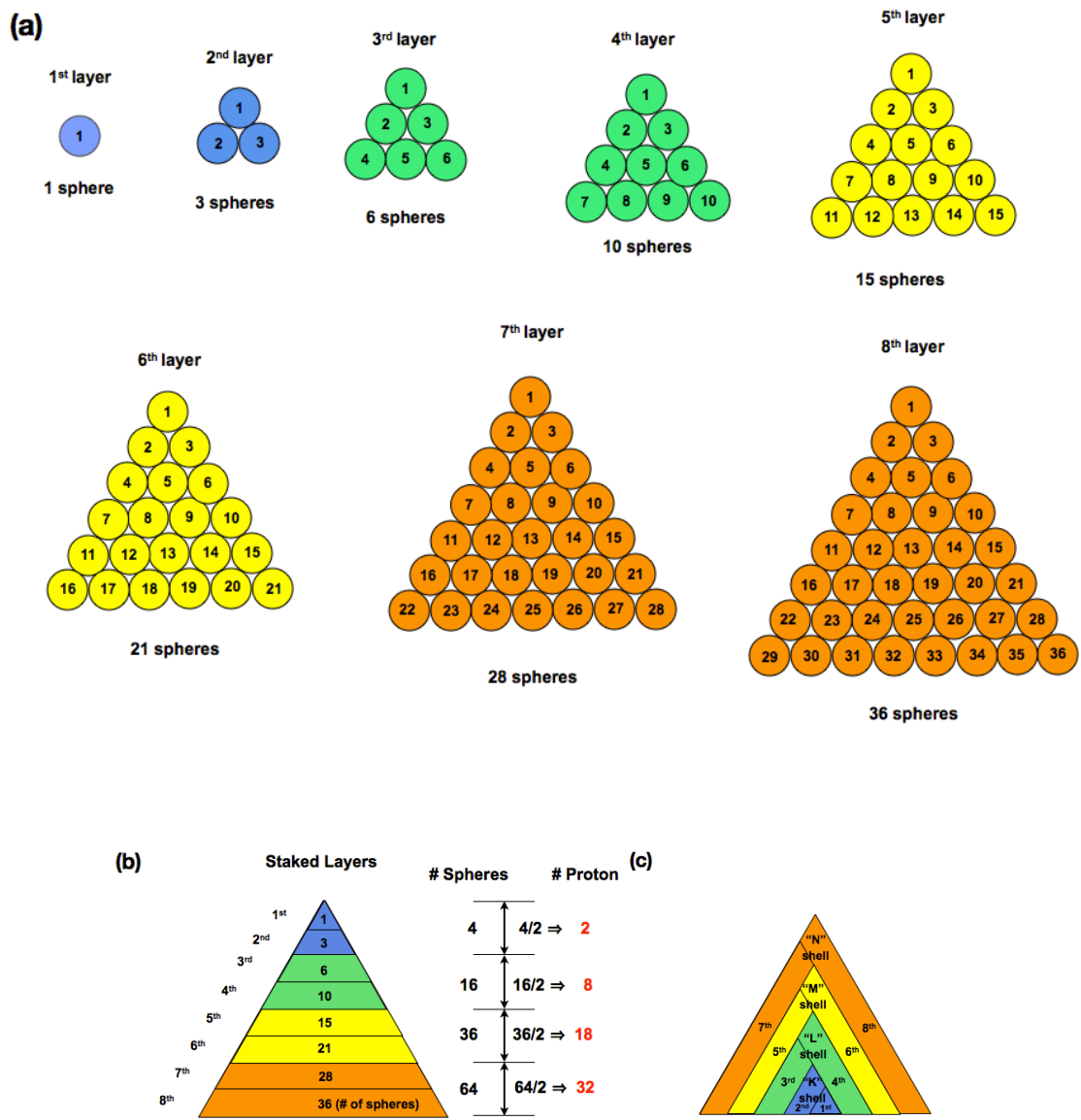

**Fig. 4.** Representing both protons and neutrons with equal spheres and arranging them in FCC structure, the number of protons in the outer layers of a tetrahedron formation is the same as the number of possible states of the principle quantum numbers.
**(a)** The number of spheres in a two dimensional closed packing arrangement in equilateral triangles.
**(b)** The number of spheres in two consecutive layers of the tetrahedron formation. Assuming a proton-neutron ratio of one, the outer layers of the tetrahedron contain the same number of protons as predicted by quantum theory.
**(c)** The same tetrahedron formation can be developed by adding the new layers to alternate sides.

The number of different positions of protons in one row of the outer shell is the same as the number of magnetic quantum numbers (Fig. 5/b). The lattice positions also reproducing the multiplicities. Thus the number of positions in an FCC lattice are identical with the quantum numbers if a tetrahedron seed is expanded. The lattice positions not only reproduce all of the quantum numbers but also bear the same physical meaning.

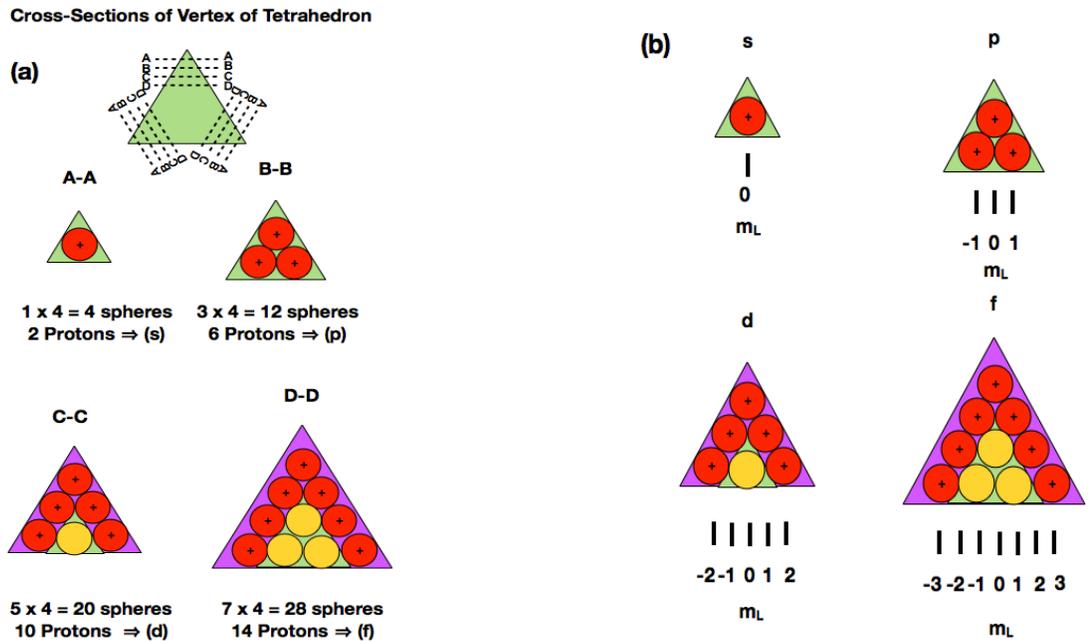

**Fig. 5.** If a tetrahedron has been developed from a core tetrahedron, which contains four spheres, then the number of protons in one layer of the outer shell of the tetrahedron is equivalent with the number of states of the angular momentum quantum number or the corresponding sub-shell. The number of different positions of the protons in one layer of the shell is the same as the number of magnetic quantum numbers. The red circles represent the outer shell nucleons. The tetrahedron has four vertexes; therefore, the number of spheres is multiplied by four.
**(a)** Number of spheres in one layer of a vertex of the tetrahedron.
**(b)** Number of the different proton positions in one layer of a vertex of a tetrahedron.

It has been assumed that higher nuclear density is preferable to lower ones. The density of structures built in FCC arrangement can be described by the ratio of the tetrahedron and octahedron sub-units. The higher ratio corresponds to higher nuclear density. It can be shown that if a tetrahedron is expanded by rotating 90 degrees at each expansion then the density of this joint or double tetrahedron is higher in comparison to a single tetrahedron. Based on the higher density it is suggested that the initial single tetrahedron should be alternately developed by rotating 90 degrees at each expansion of the tetrahedron. Three dimensional images of the completed tetrahedrons corresponding to the elements He, Ne, Ar, Kr, Xe and Ra are shown in Fig. 6/a-f. The nuclear FCC tetrahedron lattice structure expanded by rotating 90 degrees reproduces not only the quantum numbers but also the periodicity of the chemical elements.

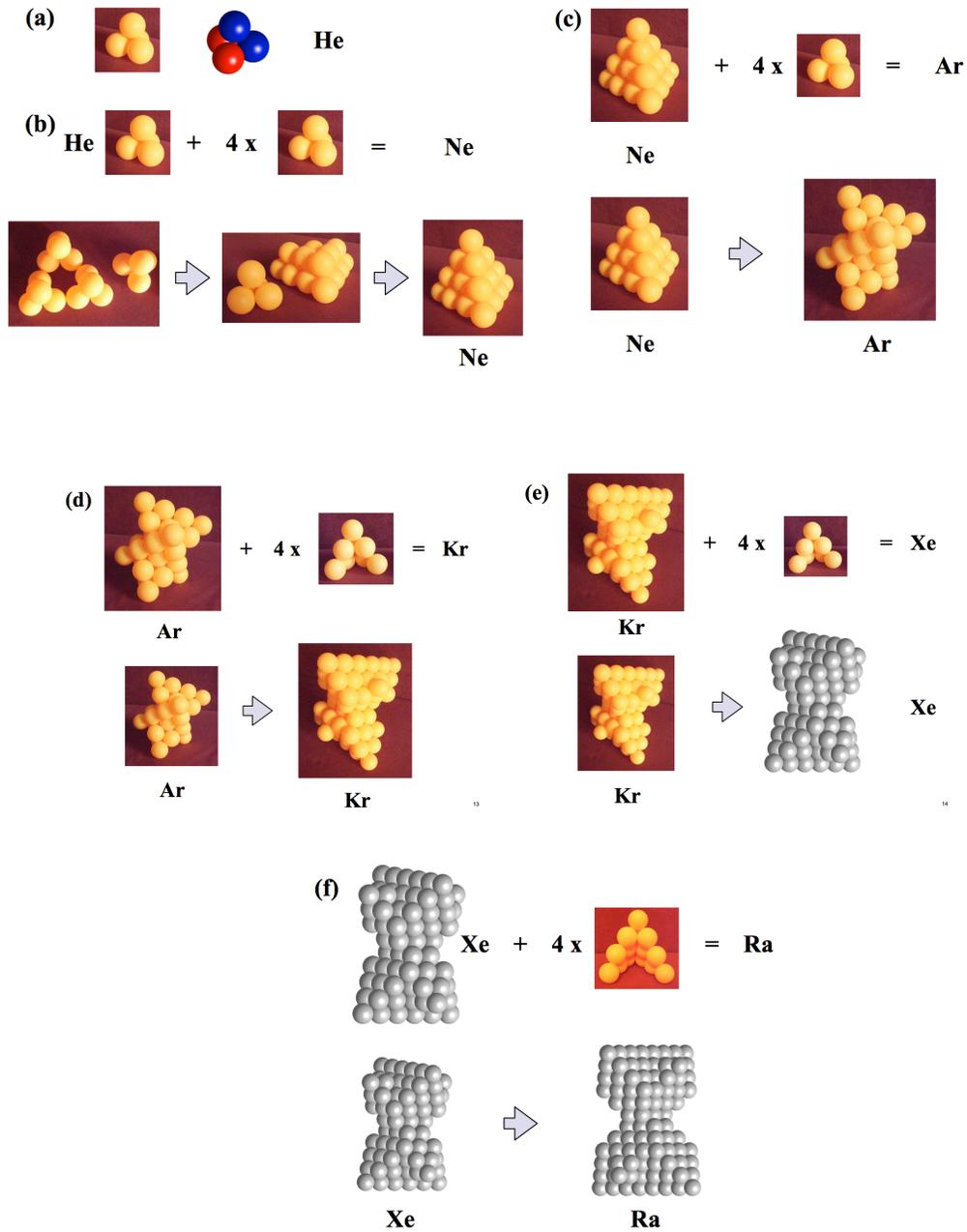

**Fig. 6.** 3D images show the completed nuclear structures of the noble gases. **(a)** Helium **(b)** Neon **(c)** Argon **(d)** Krypton **(e)** Xenon **(f)** Radon

This study focuses on the relationship between the nucleus and the PSCE; therefore, the additional characteristics of the nucleus, supporting the tetrahedron lattice model, are only listed here.

● The expansion of tetrahedron into four dimensions with angles of 109.5 degrees reproduces the original tetrahedron symmetry for every fourth nucleus. This is consistent with the observed zero magnetic momentum for each even-even nucleus.

● The disintegration of a structure should produce fragments of its basic lattice. The fragmentation of the tetrahedron FCC nuclear lattice is consistent with nuclear fission. The preferred alpha decay of the nuclear structure is consistent with the disintegration of an FCC lattice, which is built up from tetrahedron sub-units.

● The measured bulk density of the nucleus is consistent with the density of the FCC lattice arrangement (Cook and Dallacasa,1987; Cook, 2010).

● The charge density distribution of the individual elements (Hofstadler, 1961) is also consistent with the shape of the "double" tetrahedron FCC nuclear model.

## 4. Mathematical Description of the Periodic System

The proposed tetrahedron nuclear lattice model can be used to derive the sequences of the periodic table. The three noticeable sequences in the periodic table are the fundamental {1, 2, 2, 3, 3, 4, 4...}, the periodic {2, 8, 8, 18, 18, 32, 32...}, and the atomic number {2, 10, 18, 36, 54, 86, 118 ...} sequences (Fig. 1).

*4.1 Fundamental Sequence*

The relationship between the periods (n) and the sequence numbers (m) can be described as:

$$m = \frac{2n + (-1)^n + 3}{4} \quad [4]$$

*4.2 Periodic Sequence*

The number of nucleons in the $k^{th}$ layer of a tetrahedron can be calculated by the triangular number [Tr(k)] (Abramowitz & Stegun, 1964; Beiler, 1964) (Fig. 4/a)

$$Tr(k) = \frac{k}{2}(k+1). \quad [5]$$

In each structural step of its development, the tetrahedron is expanded by one layer in two directions (Fig. 4/c) giving the relationship between the tetrahedron layers and the sequence numbers as:

$$k = 2m. \quad [6]$$

The number of nucleons in the outer shell of the tetrahedron [Tr(m)] is the sum of the two consecutive triangular numbers.

$$\mathrm{Tr}(m) = \mathrm{Tr}(k) + \mathrm{Tr}(k-1) = \mathrm{Tr}(2m) + \mathrm{Tr}(2m-1) = 4m^2 \qquad [7]$$

The number of charges in the completely developed shell is

$$\Delta Z(n) = \frac{\mathrm{Tr}(m)}{2} = 2m^2 \qquad [8]$$

Substituting the sequence number from Eq. 4 gives:

$$\Delta Z(n) = \frac{1}{8}\left[2n + (-1)^n + 3\right]^2 \qquad [9]$$

which is the same formula given by Tomkeieff (1951; 1954).

### 4.3 *Atomic Number Sequence*

A formula giving the total number of charges in the nucleus with completely developed shells can be derived in a similar manner. The total number of nucleons in a tetrahedron with k layers can be determined by its tetrahedral number [Th(k)] (Abramowitz & Stegun, 1964; Conway, 1996)

$$\mathrm{Th}(k) = \frac{k}{6}(k+1)(k+2) \qquad [10]$$

Substituting the sequence number from Eq. (6) gives the number of nucleons in a tetrahedron for sequence (m) as:

$$\mathrm{Th}(m) = \frac{m}{3}(2m+1)(2m+2) = \frac{4m^3}{3} + 2m^2 + \frac{2m}{3} \qquad [11]$$

The tetrahedron nucleus is developed by alternately expanding the tetrahedrons rotated by 90 degrees (Fig. 4/d). The number of nucleons in the tetrahedron is

$$\mathrm{Th}^{double}(n) = 2\mathrm{Th}(m) - \mathrm{Tr}^{even-period}(m) - 4 \;. \qquad [12]$$

The formula

$$\frac{(-1)^n + 1}{2} \qquad [13]$$

can be used to generate 0 for odd periods and 1 for even number periods, and

$$\mathrm{Tr}^{even-period}(m) = \frac{(-1)^n + 1}{2}\mathrm{Tr}(m) \qquad [14]$$

Equation (12) can be rewritten then as

$$\mathrm{Th}^{double}(n) = 2\mathrm{Th}(m) - \frac{(-1)^n + 1}{2}\mathrm{Tr}(m) - 4 \qquad [15]$$

The number of charges in the nucleus in a completely developed sequence is

$$Z(n) = \frac{\mathrm{Th}^{double}(n)}{2}. \qquad [16]$$

Combining Eqs. (4), (7), (11), (12), (15) and (16) gives the number of nuclear charges for any period. The atomic number sequence of the periodic table can be described then as:

$$Z(n) = \frac{1}{3}\left\{4m^3 - [(-1)^n - 1]3m^2 + 2m - 6\right\}. \qquad [17]$$

Substituting m from Eq. (4) gives the atomic number sequence (Garai, 2008)

$$Z(n) = \frac{1}{48}[2n + (-1)^n + 3]^3 - \frac{1}{16}[(-1)^n - 1][2n + (-1)^n + 3]^2 + \frac{1}{6}[2n + (-1)^n + 3] - 2 \qquad [18]$$

Any physical models proposed for the explanation of PSCE has to explain and reproduce the sequences of the table.

For comparison the sequence of the Principle Quantum Numbers [PQN(n)] is given as:

$$S_{PQN} = \{2,\ 8,\ 18,\ 32...\}. \qquad [19]$$

The first two entries are the same in both the sequence of the principle quantum number and the sequence of the length of the period. The rest of the entries in these sequences shows similarity but they are not identical. The sequence of the principle quantum number can be given as:

$$PQN(n) = 2n^2. \qquad [20]$$

This formula [Eq. (20)] emerging from quantum theory does not agree with any of the sequences of the periodic table [Eqs.(4, 9, 18)] indicating that quantum mechanics can not give a viable explanation for the sequences of the periodic table.

5. **Pauli Exclusion Principle**

The chemical properties of the elements is defined by the electronic structure of the outermost or valence shell. The identity of an atom, including its electronic structure is determined by the nucleus or more specifically by the number of protons (Broek, 1911, 1913). The interaction between the atomic particles can be described as:

       Strong Force Effects      Electrostatic Effects

Neutrons      ⇔      Protons      ⇒      Electrons.

The negatively charged electrons in an atom are captured by the electrostatic attraction of the positively charged protons. The electrostatic attraction between the two differently charged particles is described by an inverse square law. The attraction between a proton and a captured electron is the function of the distance between the two charges. The energy of a captured electron in its ground state depends on the distance between the two particles. Electrons with different energies in their ground states should be separated by different distances from the protons capturing them. This distance must be invariable otherwise the electronic structure of the atom would not be stable. In order to maintain the configuration of the protons and the distances between the protons and the captured electrons the position of the protons must be invariant. This condition is fulfilled by a nuclear lattice. The lattice structure of the nucleus ensures that if a lattice is occupied then another proton can not have the same position. Thus the nuclear lattice model offers a feasible physical explanation for the Pauli Exclusion Principle, which requires that the protons in the nucleus cannot co-exist in the same location. It is concluded that in order to comply with the Pauli Exclusion Principle the protons in the nucleus should be arranged in a lattice.

## 6. Physical Explanation for the Periodicity of the Electronic Structure

The most important feature in the development of the electronic structure of the chemical elements is the {2, 8, 8, 18, 18, 32, 32...} pattern which represents the number of elements in each period of the periodic table or the number of electrons in each completed shell. The periodicity pattern for the ground state neutral atoms and for the positive atomic ions is different (Brakel, 2000; Goudsmith & Richards, 1964). The most simple manifestation of periodicity, the ground state neutral atoms, are investigated here. The periodic sequence is described by equation 9, and it is attributed to the number of protons in the outer shell of the tetrahedron nuclear structure. It will be investigated that how protons in the nuclear lattice can affect the electronic structure of the elements.

The interaction between the proton and electron is the result of the exchange of virtual photons. Based on the energies of the electrons in the atoms virtual photon or quantum electrodynamic treatment is not necessary and the classical approach is sufficient to describe the interaction. The classical approach is supported by the success of the Bohr model, which correctly describes the orbit of electron in the Hydrogen atom (Bohr, 1913).

One of the outcome of the nuclear lattice model is that the nucleus can not be considered as a point charge. The lattice position of a given proton must be taken into consideration when the attraction on an electron is calculated. The relative attractive force $[F_{e-P}(Z)]$ of a proton $[P(Z)]$ on an electron $[e(Z)]$ is characterized by the distance between the charge center of the nucleus and the proton $[d_{NCC-P(Z)}]$

$$F_{e-P}(Z) = f(d_{NCC-P(Z)}). \hspace{2cm} [21]$$

where Z is the corresponding atomic number defined as follows. The first proton in the Hydrogen nucleus is P(1), the next proton added into the structure to form the nucleus of Helium is P(2) and the last proton added to the nuclear structure of element Z-1 to form the nucleus of element Z is P(Z). The electrons are labeled accordingly. Thus electron(1) is captured by P(1) and electron(Z) is captured by P(Z). Depending on the lattice positions of proton(Z) the attraction can be either stronger or weaker than the attraction of proton(Z-1). The relative attraction of proton Z-1 and Z can be defined by comparing the distances between the charge center of the nucleus and the position of the proton:

$$d_{P-NCC}(Z-1) < d_{P-NCC}(Z) \quad \text{then} \quad F_{P-e}(Z-1) < F_{P-e}(Z) \hspace{2cm} [22]$$

or vice versa. In order to overcome the repulsion force of the existing valence electrons, the relative attractive force on the captured electron(Z) must be stronger than the relative attraction force on the electrons already occupying the valence shell. Thus the following inequality must be satisfied:

$$F_{P-e}(Z) > F_{P-e}(Z-1) \quad \Leftrightarrow \quad d_{P-NCC}(Z) > d_{P-NCC}(Z-1). \hspace{2cm} [23]$$

This inequality is consistent with the Aufbau principle, which states that orbitals with the lowest energy are filled first. When this condition [Eq. 23] is not satisfied then it is predicted that the electron(Z) is unable to join to the valence shell and starts to form a new valence shell outside of the existing one. Thus a new electron shell starts to form when the condition [Eq. 24] is satisfied.

$$F_{P-e}(Z) < F_{P-e}(Z-1) \quad \Leftrightarrow \quad d_{P-NCC}(Z) < d_{P-NCC}(Z-1) \hspace{2cm} [24]$$

Based on stability considerations the first proton in the new nuclear shell should be positioned at the middle of the face of the tetrahedron and this new outer layer of the structure should develop towards the edges. It can be seen by visual inspection that in a given nuclear shell equation 23 or the Aufbau principle is satisfied (Fig. 7/a).

The distance of the last proton completing a shell [n] and the distance of the first proton in the new shell [n+1] is calculated from the nuclear charge center. It is assumed that the position of the last proton in layer n is at the vertex of the tetrahedron and the position of the first proton in layer n+1 is at the surface, closest to the charge center (Fig. 7/b). It is also assumed that the charge center of the nucleus coincides with the mass center. For the completed structures of the tetrahedron He, Ne and "double" tetrahedron Ar, Xe and Ra nucleus, the mass center coincides with the charge centers. When a new nuclear shell is started by adding a proton, then the charge center shifts towards the proton resulting in smaller distance between the charge center and the proton than the distance between the proton and the mass center. Thus using the mass center instead of the charge center is a conservative estimate on the inequality of $d_{P-NCC}(Z) < d_{P-NCC}(Z-1)$.

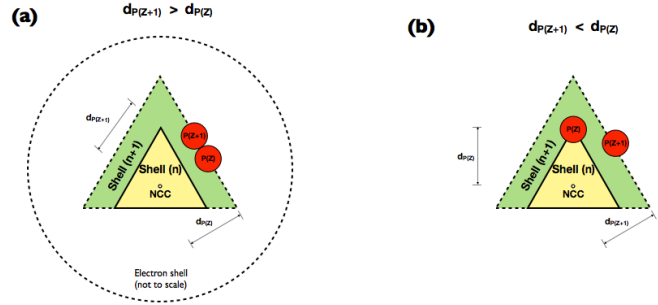

**Fig. 7.** Schematic figure in 2D showing the geometry of the nucleus.
**(a)** The distance between the the nuclear charge center (NCC) and the proton increases as the protons occupy the lattice positions in the same layer.
**(b)** When a new layer starts to build up in the nucleus, the proton is closer towards to the nuclear charge center than the proton completing the previous shell. Thus proton [P(Z+1)] starting the period of n+1 has weaker attraction on the captured electron than the proton [P(Z)] completing the period n has. The captured electron(Z+1) is unable to overcome the repulsion of the existing shell because of its weaker attraction and starts to built up a new valence shell outside of the existing one.

The distance between the first proton in the shell and the mass center is calculated first. For simplification the length of the edge of the basic tetrahedron is one unit. Thus the diameter of the nucleons is also one unit. The height of a tetrahedron [$h_\Delta$] with unit length is

$$h_\Delta = \sqrt{\frac{2}{3}} \, . \quad [25]$$

Assuming a unit mass for the nucleons the distance between the vertex of the tetrahedron and the mass center is:

$$d_{\Delta vertex - MC} = \frac{3}{4} h_\Delta = \sqrt{\frac{3}{8}} \quad [26]$$

The distance between the face of the tetrahedron and the mass center is then:

$$d_{\Delta surface - MC} = \frac{1}{4} h_\Delta = \sqrt{\frac{1}{24}} \quad [27]$$

The tetrahedron is expanded by one-one layers on both sides; thus the length of the side of the tetrahedron is increased by two units in each periods (Fig. 7). The distance between the center of a sphere (nucleon lattice position), placed on the surface of the tetrahedron, and the mass center in period n+1 is then

$$d_{\Delta_{surface}-MC}(n+1) = (2n+1)\frac{1}{4}h_\Delta = (2n+1)\sqrt{\frac{1}{24}} \qquad [28]$$

where n is the number of period. The distance between a sphere placed at the vertex of the tetrahedron in period n can be calculated then as:

$$d_{\Delta_{vertex}-MC}(n) = (2n-1)\frac{3}{4}h_\Delta = (2n-1)\sqrt{\frac{3}{8}} \qquad [29]$$

It can be seen that

$$d_{\Delta vertex-MC}(n) > d_{\Delta surface-MC}(n+1) \qquad [30]$$

where n = 2, 3, 4, 5...  Arranging the nucleons in an FCC lattice and building up a tetrahedron inequality in equation 24 is fullfilled when a new layer in the nuclear structure starts to form (Fig. 7/b). Thus

$$\begin{aligned} &d_{P-NCC}(Z=2) > d_{P-NCC}(Z=3) \bullet d_{P-NCC}(Z=10) > d_{P-NCC}(Z=11) \bullet d_{P-NCC}(Z=18) > d_{P-NCC}(Z=19) \\ &d_{P-NCC}(Z=36) > d_{P-NCC}(Z=37) \bullet d_{P-NCC}(Z=54) > d_{P-NCC}(Z=55) \bullet d_{P-NCC}(Z=86) > d_{P-NCC}(Z=87). \end{aligned} \qquad [31]$$

The remaining rest of the pairs of the elements ($[Z] \Leftrightarrow [Z+1]$) satisfy the relationship of equation 23 and $d_{P-NCC}(Z) > d_{P-NCC}(Z-1)$.

Resulting from the inequalities of Eq. 31 the relative attraction force on the newly captured electron is weaker than the force on the electrons in the existing shell leads to the formation of a new electron shell for the following elements:

$$\begin{aligned} &F_{P-e}(He) > F_{P-e}(Li) \bullet F_{P-e}(Ne) > F_{P-e}(Na) \bullet F_{P-e}(Ar) > F_{P-e}(K) \\ &F_{P-e}(Kr) > F_{P-e}(Rb) \bullet F_{P-e}(Xe) > F_{P-e}(Cs) \bullet F_{P-e}(Rn) > F_{P-e}(Fr) \end{aligned} \qquad [32]$$

Based on the presented geometrical consideration it is suggested that the cyclical structural development of the nuclear structure results in an interruption of the development of valence shell because the relative attraction of the proton on the captured electron becomes weaker when a new layer starts to built in the nucleus (Garai, 2011). The cycles of the weaker relative attractions predicted from the geometry of the nuclear structure are identical with the length of the periods in the periodic table. It is concluded that the electronic configuration of the chemical elements is the consequence of the structural development of the nuclear lattice.

## 7. Conclusions

The symmetry pattern of the periodic system of the chemical elements emerges from the nuclear charge. This invariant pattern can be maintained only if the positions of the nucleons are also invariant. Thus the nucleons should form a lattice structure. The lattice arrangement of the nucleons is consistent with the Pauli exclusion principle and offers a feasible physical explanation for the exclusion principle.  Representing protons and neutrons with equal spheres, arranging them alternately in an FCC lattice, and developing a tetrahedron, rotating by 90 degrees at each expansion results in an identical symmetry pattern expressed by the PSCE. The lattice positions of the nucleons not only reproduces the quantum numbers but they also bear the same physical meaning indicating that the presented model should be considered as a credible physical explanation for quantum theory.

Based on the presented nuclear lattice model mathematical solutions for the sequences of the periodic table are derived. Investigating the structural development of the nucleus it is shown that the periodicity of the electronic structure is the natural outcome of the nuclear lattice geometry.  The presented tetrahedron lattice model is the first nucleus model which is able to reproduces all of the sequences of the periodic table, offers a credible physical explanation for the identical symmetry of the nucleus and the electronic structure, for the Pauli exclusion principle and for the Aufbau principle.